\documentclass[apj]{emulateapj}
\usepackage{epsf}
\usepackage{amsmath,graphicx}
\usepackage[z]{tmsfnts}
\def\fgas{\ensuremath{f_{\rm gas}}}
\def\Mgas{\ensuremath{M_{\rm gas}}}
\def\Mtot{\ensuremath{M_{\rm tot}}}

\def\Tave{\ensuremath{\langle T \rangle}}
\def\Tspec{\ensuremath{T_{\rm spec}}}
\def\TspecIV{\ensuremath{T_{\rm spec}[0.5\r500c]}}
\def\Tmg{\ensuremath{T_{\text{mg}}}}
\def\TmgSIM{\ensuremath{T_{\text{mg}}^{\text{SIM}}}}

\def\Tew{\ensuremath{T_{\text{ew}}}}

\def\Msun{\ensuremath{M_\odot}}

\def\R2500c{\ensuremath{r_{\text{2500c}}}}
\def\r500c{\ensuremath{r_{\text{500c}}}}
\def\M2500c{\ensuremath{M_{\text{2500c}}}}
\def\m500c{\ensuremath{M_{\text{500c}}}}

\def\M500{\ensuremath{M_{\text{500}}}}
\def\T500{\ensuremath{T_{\text{500}}}}
\def\P500{\ensuremath{P_{\text{500}}}}
\def\K500{\ensuremath{K_{\text{500}}}}
\def\***#1{\textbf{\textsf{***#1***}}}

\shorttitle{TESTING X-RAY CLUSTER MEASUREMENTS WITH SIMULATIONS}
\shortauthors{NAGAI, VIKHLININ \& KRAVTSOV}

\begin{document}

\submitted{The Astrophysical Journal, submitted}
\slugcomment{The Astrophysical Journal, submitted} 

\title{Testing X-Ray Measurements of Galaxy Clusters with Cosmological
  Simulations}

\author{%
Daisuke~Nagai\altaffilmark{1},
Alexey~Vikhlinin\altaffilmark{2,3},
Andrey~V.~Kra\kern-0.05emvtsov\altaffilmark{4}
}

\begin{abstract}   
  X-ray observations of galaxy clusters potentially provide powerful
  cosmological probes if systematics due to our incomplete knowledge
  of the intracluster medium (ICM) physics are understood and
  controlled. In this paper, we present mock {\it Chandra} analyses of
  cosmological cluster simulations and assess X-ray measurements of
  galaxy cluster properties using a model and procedure essentially
  identical to that used in real data analysis.  We show that
  reconstruction of three-dimensional ICM density and temperature
  profiles is excellent for relaxed clusters, but still reasonably
  accurate for unrelaxed systems.  The total ICM mass is measured
  quite accurately ($\lesssim$6\%) in all clusters, while the
  hydrostatic estimate of the gravitationally bound mass is biased low
  by about 5\%--20\% through the virial region, primarily due to
  additional pressure support provided by subsonic bulk motions in the
  ICM, ubiquitous in our simulations even in relaxed systems. Gas
  fraction determinations are therefore biased high; the bias
  increases toward cluster outskirts and depends sensitively on its
  dynamical state, but we do not observe significant trends of the
  bias with cluster mass or redshift. We also find that different
  average ICM temperatures, such as the X-ray spectroscopic $\Tspec$
  and gas-mass-weighted $\Tmg$, are related to each other by a
  constant factor with a relatively small object-to-object scatter and
  no systematic trend with mass, redshift or the dynamical state of
  clusters. We briefly discuss direct applications of our results for
  different cluster-based cosmological tests.
\end{abstract}



\altaffiltext{1}{Theoretical Astrophysics, California Institute of
Technology, Mail Code 130-33, Pasadena, CA 91125 ({\tt daisuke@caltech.edu})}
\altaffiltext{2}{Harvard-Smithsonian Center for Astrophysics, 60 Garden Street, Cambridge, MA 02138}
\altaffiltext{3}{Space Research Institute, 8432 Profsojuznaya St., GSP-7, Moscow 117997, Russia}
\altaffiltext{4}{Department of Astronomy and
Astrophysics, Kavli Institute for Cosmological Physics, The Enrico Fermi Institute, 5640 South
Ellis Ave., The University of Chicago, Chicago, IL 60637}

\section{Introduction}
\label{sec:intro}

X-ray observations of clusters of galaxies can potentially provide plethora of
useful cosmological information. The cluster-based cosmological tests include
cluster number counts based on the temperature function, baryon fraction and
X-ray and SZ Hubble constant measurements.  These cosmological tests provide
potentially very powerful constraints on the matter density, $\Omega_M$,
$\Omega_{\Lambda}$, $\sigma_8$, and the equation of state of dark energy, $w$.
In all of these cosmological tests, the key observational ingredients are the
gas mass ($\Mgas$), total mass ($\Mtot$), gas mass fractions ($\fgas\!
\equiv\!  \Mgas/\Mtot$), and average cluster temperature ($\Tave$).  In the
era of precision cosmology, it is paramount to achieve accurate measurements
of the key cluster parameters, check how they depend on various simplifying
assumptions, and control systematics due to our incomplete knowledge of the
ICM physics.

There are many important simplifying assumptions used in deriving
cluster properties from X-ray data. For example, it is usually assumed
that the ICM density is a function of radius only and does not have
small-scale substructure. However, substructure in clusters is
ubiquitous and this biases the X-ray derived $\Mgas$ high
\citep{mathiesen_etal99}. Also, the small-scale clumps are often
associated with subhalos and thus have lower temperature. Neglecting
this can bias the average spectroscopic temperature of the cluster as
a whole \citep{mathiesen_etal01,mazzotta_etal04} which could
potentially lead to biases in constraints on cosmological parameters
\citep{rasia_etal05}. Cluster total masses are often estimated
assuming the hydrostatic equilibrium for the ICM.  This may
underestimate the true gravitational bound mass if non-thermal
pressure support is present.  For example, turbulent gas motions can
provide about 10\%--20\% of the total pressure support even in relaxed
clusters and hence bias the hydrostatic estimate
\citep{schuecker_etal04,faltenbacher_etal05,rasia_etal06,lau_etal06}.
Spherical symmetry is another commonly used assumption useful in
solving for three-dimensional (3D) physical cluster properties from
two-dimensional (2D) observed quantities, but could bias results if
clusters are aspherical.

X-ray observations with {\it Chandra} and {\it XMM-Newton } enable us
to study properties of the ICM with unprecedented detail and accuracy
and provide important handles on the ICM modeling and associated
systematics. Their superb spatial resolution and sensitivity enable
accurate X-ray brightness and temperature measurements at a large
fraction of the cluster virial radius and also make it simple to
detect most of the small-scale X-ray clumps. Despite this recent
observational progress, the biases in the determination of the key
cluster properties remain relatively uncertain. The main obstacle is
that because of their unrivaled statistical accuracy, the X-ray
results cannot be contrasted against any other independent
observational techniques. In this study, we attempt to check the
validity of the X-ray analyses, with a specific focus on the analyses
of high-resolution {\it Chandra} data, by using mock observations of
clusters derived from cosmological simulations, for which the true
answers are known.

Our simulations properly treat both collisionless dynamics of dark
matter and stars and gasdynamics and capture a variety of physical
phenomena from the nonlinear collapse and merging of dark matter to
shock-heating and radiative cooling of gas, star formation, chemical
enrichment of the ICM by supernova and energy feedback. These
simulations can therefore be used to test observational biases due to
incomplete relaxation of gas, the dynamical state of a cluster,
substructure, or non-isothermality. As we show in the forthcoming
papers, the current simulations match the observed ICM profiles
outside cluster cores \citep{nagai_etal06c} and the global gas mass 
fraction \citep {kravtsov_etal06b} and are therefore sufficiently 
realistic for a purpose of the current study.

The mock observations of the simulated clusters are generated by
exactly reproducing the response properties of the \emph{Chandra}
telescope. The analysis of the mock data is identical to that used for
\emph{real Chandra} observations by
\cite{vikhlinin_etal05c,vikhlinin_etal06}.  The comparison of the true
and derived cluster properties provides an assessment of any biases
introduced by the X-ray analysis. Some of these biases are potentially
redshift-dependent (e.g, those related to higher merger rate or
decreased ability to detect small-scale substructures in the X-ray
data for high-$z$ systems). To check for any redshift dependence in
such biases we use the simulation outputs at $z=0$ and $0.6$. Our
results indicate that the X-ray analysis provides accurate
reconstruction of the 3D properties of the ICM. The strongest biases
we find are those in the hydrostatic mass estimates. They are related
to physics explicitly missing from the hydrostatic method (e.g.,
turbulence), and not to deficiencies of the X-ray analysis.

\section{Cosmological Cluster Simulations}
\label{sec:sim}

\label{sec:numdetails}

In this study, we analyze high-resolution cosmological simulations of
16 cluster-sized systems in the flat {$\Lambda$}CDM model:
$\Omega_{\rm m}=1-\Omega_{\Lambda}=0.3$, $\Omega_{\rm b}=0.04286$,
$h=0.7$ and $\sigma_8=0.9$, where the Hubble constant is defined as
$100h{\ \rm km\ s^{-1}\ Mpc^{-1}}$, and an $\sigma_8$ is the power
spectrum normalization on an $8h^{-1}$~Mpc scale.  The simulations
were done with the Adaptive Refinement Tree (ART)
$N$-body$+$gasdynamics code \citep{kravtsov99, kravtsov_etal02}, a
Eulerian code that uses adaptive refinement in space and time, and
(non-adaptive) refinement in mass \citep{klypin_etal01} to reach the
high dynamic range required to resolve cores of halos formed in
self-consistent cosmological simulations. The simulations presented
here are discussed in detail in \citet{kravtsov_etal06} and
\citet{nagai_etal06c} and we refer the reader to these papers for more
details. Here we summarize the main parameters of the simulations.

High-resolution simulations were run using a uniform 128$^3$ grid and 8
levels of mesh refinement in the computational boxes of
$120\,h^{-1}$~Mpc for CL101--107 and $80\,h^{-1}$~Mpc for CL3--24 (see
Table~\ref{tab:sim}). These simulations achieve a dynamic range of
$32768$ and a formal peak resolution of $\approx 3.66\,h^{-1}$~kpc and
$2.44\,h^{-1}$~kpc, corresponding to the actual resolution of $\approx
7\,h^{-1}$~kpc and $5\,h^{-1}$~kpc for the 120 and $80\,h^{-1}$~Mpc
boxes, respectively.  Only the region of $\sim 3-10\,h^{-1}$~Mpc around
the cluster was adaptively refined, the rest of the volume was followed
on the uniform $128^3$ grid.  The mass resolution, $m_{\rm part}$,
corresponds to the effective $512^3$ particles in the entire box, or the
Nyquist wavelength of $\lambda_{\rm Ny}=0.469\,h^{-1}$ and
$0.312\,h^{-1}$ comoving Mpc for CL101--107 and CL3--24, respectively,
or $0.018\,h^{-1}$ and $0.006\,h^{-1}$~Mpc in the physical units at the
initial redshift of the simulations. The dark matter particle mass in
the region around the cluster was $9.1\times 10^{8}\,h^{-1}\, M_{\odot}$
for CL101--107 and $2.7\times 10^{8}\,h^{-1}\,M_{\odot}$ for CL3--24,
while other regions were simulated with lower mass resolution.

The $N-$body$+$gasdynamics cluster simulations used in this analysis
include collisionless dynamics of dark matter and stars, gasdynamics
and several physical processes critical to various aspects of galaxy
formation: star formation, metal enrichment and thermal feedback due
to supernovae Type II and Type Ia, self-consistent advection of
metals, metallicity dependent radiative cooling and UV heating due to
cosmological ionizing background \citep{haardt_madau96}. The cooling
and heating rates take into account Compton heating and cooling of
plasma, UV heating, and atomic and molecular cooling, and are
tabulated for the temperature range $10^2<T<10^9$~K and a grid of
metallicities, and UV intensities using the {\tt Cloudy} code
\citep[ver.  96b4;][]{ferland_etal98}. The Cloudy cooling and heating
rates take into account metallicity of the gas, which is calculated
self-consistently in the simulation, so that the local cooling rates
depend on the local metallicity of the gas. Star formation in these
simulations was done using the observationally-motivated recipe
\citep[e.g.,][]{kennicutt98}: $\dot{\rho}_{\ast}=\rho_{\rm
  gas}^{1.5}/t_{\ast}$, with $t_{\ast}=4\times 10^9$~yrs. The code
also accounts for the stellar feedback on the surrounding gas,
including injection of energy and heavy elements (metals) via stellar
winds, supernovae, and secular mass loss. 

These simulations therefore follow the formation of galaxy clusters
starting from the well-defined cosmological initial conditions and
capture the dynamics and properties of the ICM in a realistic
cosmological context.  However, some potentially relevant physical
processes, such as AGN bubbles, magnetic field, and cosmic rays, are
not included.  Therefore, the simulated ICM properties are probably
not fully realistic in the innermost cluster regions. Moreover, the
gas in the simulations is treated as an ideal inviscid fluid with a
small amount of numerical viscosity, and it remains unclear to what
extent the level of ICM turbulence found in the simulations and
discussed below applies to real clusters.  Despite these limitations,
the current simulations reproduce the observed ICM profiles outside
cluster cores \citep{nagai_etal06c} and the global gas mass fraction 
\citep {kravtsov_etal06b} and are therefore sufficiently realistic 
for a purpose of the current study. 

%
%
%
%
%
%

\begin{deluxetable}{lccccc}
\tablecolumns{6}
\tablecaption{Simulated cluster sample of the CSF run at $z$=0\label{tab:sim}}
\tablehead{
\multicolumn{1}{c}{Name\hspace*{8mm}}&
\multicolumn{1}{c}{}&
\multicolumn{1}{c}{$\r500c$} &
\multicolumn{1}{c}{$\m500c^{\rm gas}$} &
\multicolumn{1}{c}{$\m500c^{\rm tot}$} &
\multicolumn{1}{c}{$\langle \Tspec \rangle$\tablenotemark{a}} 
\\ 
\multicolumn{2}{c}{}&
\multicolumn{1}{c}{($h^{-1}$Mpc)}&
\multicolumn{1}{c}{($h^{-1}10^{13}\Msun$)}& 
\multicolumn{1}{c}{($h^{-1}10^{14}\Msun$)} &
\multicolumn{1}{c}{(keV)} 
}
\startdata
CL101 \dotfill & & 1.160 & 8.19 & 9.08 & 8.7 \\
CL102 \dotfill & & 0.978 & 4.83 & 5.45 & 5.6 \\
CL103 \dotfill & & 0.993 & 4.93 & 5.71 & 4.7 \\
CL104 \dotfill & & 0.976 & 5.17 & 5.39 & 7.7 \\
CL105 \dotfill & & 0.943 & 4.73 & 4.86 & 6.2 \\
CL106 \dotfill & & 0.842 & 3.18 & 3.47 & 4.3 \\
CL107 \dotfill & & 0.762 & 2.17 & 2.57 & 4.0 \\
CL3   \dotfill & & 0.711 & 1.92 & 2.09 & 3.7 \\
CL5   \dotfill & & 0.609 & 1.07 & 1.31 & 1.9 \\
CL6   \dotfill & & 0.661 & 1.38 & 1.68 & 2.0 \\
CL7   \dotfill & & 0.624 & 1.22 & 1.41 & 1.9 \\
CL9   \dotfill & & 0.522 & 0.74 & 0.82 & 1.1 \\
CL10  \dotfill & & 0.487 & 0.43 & 0.67 & 2.4 \\
CL11  \dotfill & & 0.537 & 0.78 & 0.90 & 3.4 \\
CL14  \dotfill & & 0.509 & 0.62 & 0.77 & 3.0 \\
CL24  \dotfill & & 0.391 & 0.26 & 0.35 & 1.5 
\enddata
\tablenotetext{a}{Spectral temperature measured in the shell of
  [0.15,1]$r_{500c}$.} 
\end{deluxetable}

\begin{figure*}
  \centerline{%
  \includegraphics[width=1.0\columnwidth]{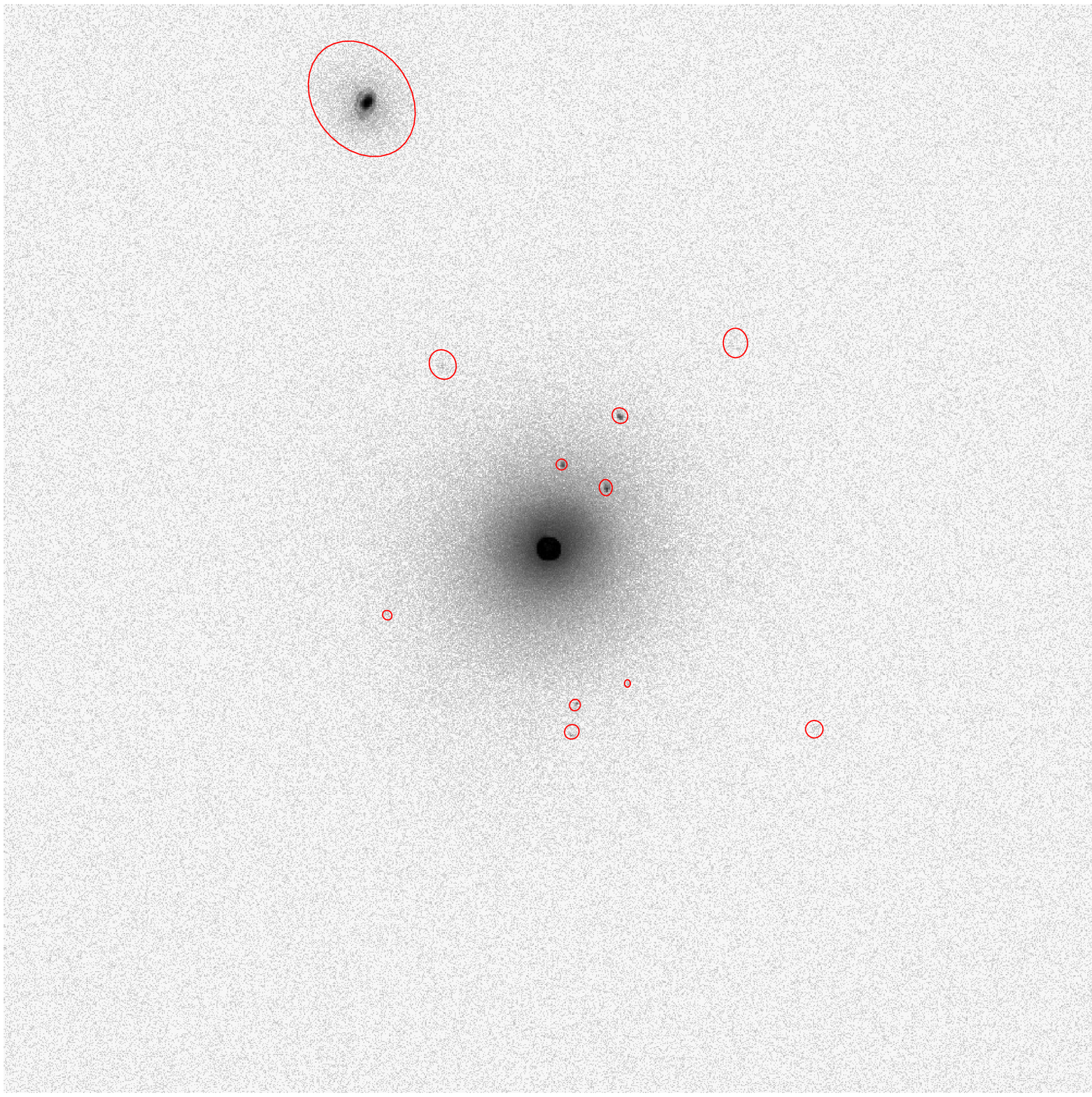}\hfill%
  \includegraphics[width=1.0\columnwidth]{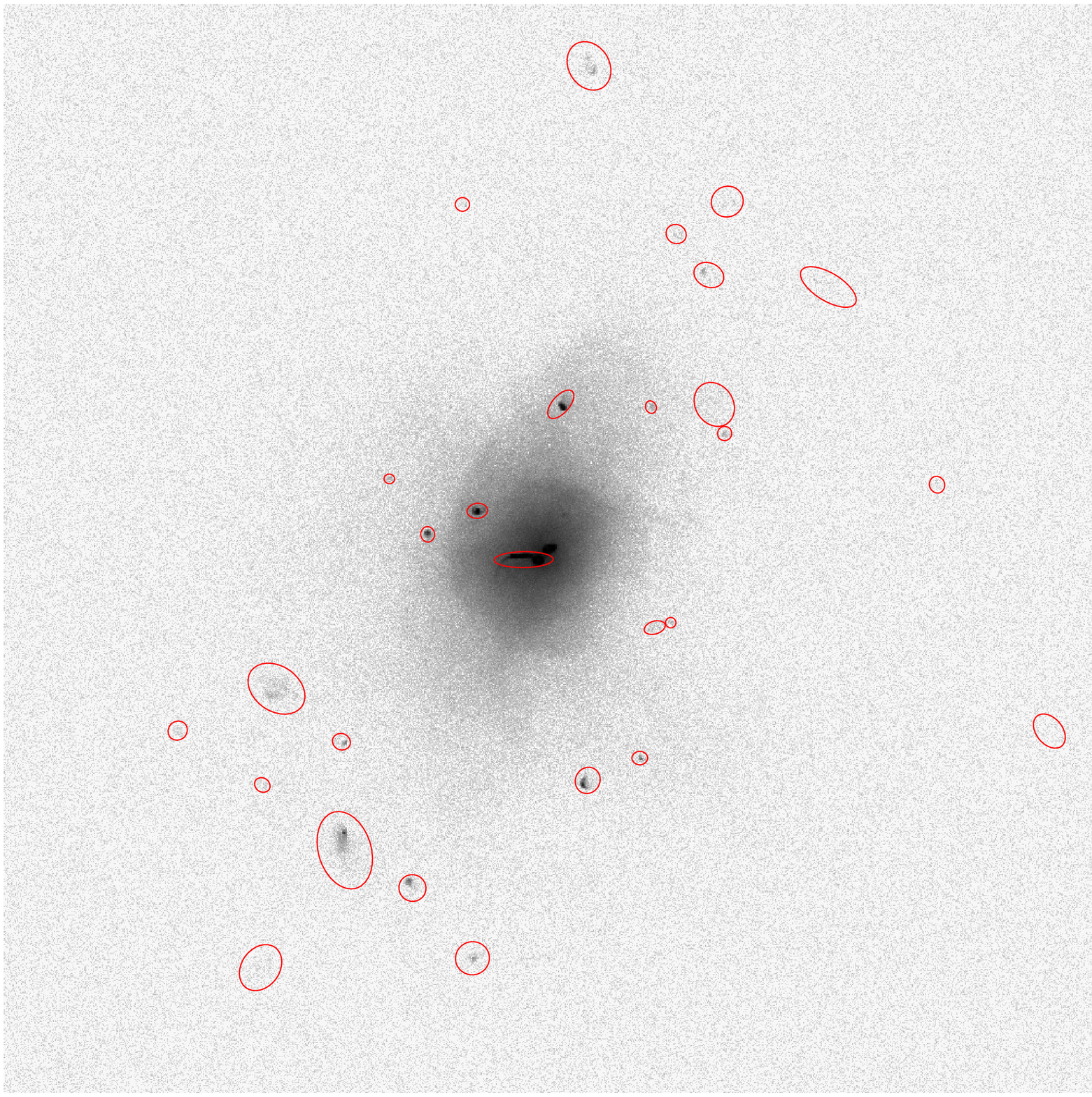}}
\caption{ Mock \emph{Chandra} images of the $z=0$ simulated clusters. 
  The \emph{left} and \emph{right} panels show the images of the relaxed
  cluster CL104 and unrelaxed CL101, respectively.  The detectable
  extended X-ray sources (other than the cluster itself) are detected
  and masked out from the analysis (red ellipses).  The physical size of
  the images is $5\,h^{-1}$~Mpc. }
\label{fig:images_z0}
\end{figure*}

\label{sec:Sample}

Our simulated sample includes 16 clusters at $z=0$ and their most
massive progenitors at $z=0.6$. The properties of simulated clusters
at $z=0$ are given in Table~\ref{tab:sim}. The masses are reported at
the radius $\r500c$ enclosing overdensities with respect to the
critical density at the redshift of the output (below, we also use a
higher overdensity level, 2500).

The mass measurement biases obviously need to be considered separately
for dynamically relaxed and non-relaxed clusters. Our relaxed
subsample is identified based on the overall structural morphology of
their \emph{Chandra} images, mimicking the procedure used by
observers.  Specifically, we visually examine mock $100$~ksec images
and identify ``relaxed'' clusters as those with regular X-ray
morphology and no secondary maxima and minimal deviations from
elliptical symmetry.  By contrast, ``unrelaxed'' clusters are those
with secondary maxima, filamentary X-ray structures, or significant
isophotal centroid shifts.  The typical examples of systems classified
as relaxed or unrelaxed are shown in Figure~\ref{fig:images_z0}.

\section{Mock \emph{Chandra} Analyses}
\label{sec:MockChandra}

\subsection{Simulating the Mock Chandra Data}
\label{sec:MockChandra1}

We create X-ray flux maps of the simulated clusters viewed along three
orthogonal projections.  The flux map is computed by projecting X-ray
emission arising from hydrodynamic cells enclosed within $3\,R_{\rm vir}$ of
a cluster along a given line-of-sight.  We compute the X-ray spectrum
arising from the $i$-th hydrodynamical cell with a cell volume $V_{i}$ as
\begin{equation}
  j_{E,i} = n_{e,i}\, n_{p,i}\, \Lambda_{E}(T_i,Z_i,z) V_{i},
\label{eq:xray}
\end{equation}
where $n_{e,i}$, $n_{p,i}$, $T_i$, and $Z_i$ are the electron and
proton densities, gas temperature, and metal abundance of the
hydrodynamic cell. We compute the X-ray plasma emissivity,
$\Lambda_{E}(T_i,Z_i,z)$, using the \textsc{mekal} code
\citep{1985A&AS...62..197M,1993A&AS...97..443K,1995ApJ...438L.115L}
with the relative heavy element abundance from \citet{anders_etal89}. 
We do not compute emission from the gas with $T<10^6$~K (0.086~keV)
because it is below the lower limit of the \emph{Chandra} bandpass. 
The plasma spectrum is multiplied by the photoelectric absorption
corresponding to the hydrogen column density of
$n_H=2\times10^{20}$~cm$^{-2}$. The resulting map consists of
$1024\times1024$ 2D image at 218 energy bins ranging from 0.1 to
10keV.  The pixel size is $4.88\,h^{-1}$~kpc for CL101--CL107 and
$2.44\,h^{-1}$~kpc for CL3--CL24.  The map size is therefore
$5\,h^{-1}$~Mpc and $2.5\,h^{-1}$~Mpc, respectively.  Throughout this
paper, we assume the cluster redshift of $z_{\rm obs}=0.06$ for the
$z=0$ sample and $z_{\rm obs}=0.6$ for the $z=0.6$ sample. 

Next, we simulate mock \emph{Chandra} data. We convolve the emission
spectrum with the response of the \emph{Chandra} front-illuminated
CCDs and draw a number of photons at each position and spectral
channel from the Poisson distribution. We simulate two sets of mock
\emph{Chandra} data.  The first set has an exposure time of 100~ksec
(typical for deep observations) and includes a background with the
intensity corresponding to the quiescent background level in the
ACIS-I observations \citep{markevitch_etal03}.  From these data, we
generate images in the 0.7--2~keV band and use them to identify and
mask out from the further analysis all detectable small-scale clumps,
as routinely done by observers.  Our clump detection is fully
automatic and based on the wavelet decomposition algorithm described
in \citet{vikhlinin_etal98}. Detection thresholds were chosen to allow
3--4 false detections in each image.  The typical limiting flux for
detection of compact extended sources is $\sim
3\times10^{-15}$~erg~s$^{-1}$~cm$^{-2}$ in the 0.5--2~keV band; this
corresponds to a luminosity of $\sim 1.5\times10^{42}$~erg~$s^{-1}$
for clusters at $z_{\rm obs}=0.06$. Detected clumps are indicated by
red ellipses in the images presented in Fig.\ref{fig:images_z0}.  Note
that the $\Mgas$ associated with the excluded clumps is no more than a
few percent of the total enclosed $\Mgas$ within \r500c at $z=0.06$. 
At $z=0.6$, considerably smaller numbers of clumps are
detected\footnote{Only a few clumps are detected in massive clusters
  with $\Tspec > 3$~keV and none in less massive systems.}, and the
clump removal has very little effect on the analysis of high-$z$
clusters. 

Once the clump detection is done on the first set, all further
analysis is performed using a second set of photon maps generated for
very long exposures yielding $\approx 10^6-10^7$ photons outside a
cluster core region in each of the simulated clusters. The exposures
are artificially long by design as we are interested in exploring
intrinsic limitations of the X-ray analysis, not the statistical
errors due to Poisson noise in a particular choice of short exposure. 
Also, we ignore further complications present in reduction of real
\emph{Chandra} data, including background subtraction and spatial
variations of the effective area (i.e., we assume that accurate
corrections for these effects can be applied to the real data and any
associated uncertainties are included in the reported measurement
errors).

\subsection{Analysis of Mock Chandra Data}
\label{sec:MockChandra2}

Using the simulated mock data as an input, we repeat the analysis
applied by \citet{vikhlinin_etal06} and \citet{kotov_vikhlinin_06} to
real \emph{Chandra} observations of relaxed clusters at $z\approx0$
and $0.5$. The purpose of this analysis is to reconstruct spherically
averaged ICM density and temperature profiles, and to estimate total
cluster mass assuming hydrostatic equilibrium.  The entire procedure
and justification of the choices made is extensively discussed in
\citet{vikhlinin_etal06}. Below, we briefly outline the essential
steps. 

The main observational input is azimuthally averaged X-ray surface
brightness and projected gas temperature and metallicity profiles. 
The temperature and metallicity profiles are obtained from a
single-temperature fit to the spectra in the 0.5--10~keV extracted in
annuli with $r_{\text{out}}/r_{\text{in}}=1.25$ centered on a main
cluster X-ray peak. 

The surface brightness profiles are extracted in the 0.7--2~keV band in
narrow concentric annuli ($r_{\text{out}}/r_{\text{in}}=1.1$). Using the
effective area as a function of energy and observed projected
temperature and metallicity at each radius, we then convert the observed
\emph{Chandra} count rate in the 0.7--2~keV band into the projected
emission measure, $\mathit{EMM}=\int n_e \,n_p\,dl$. 

Three-dimensional profiles of $\rho_{\rm gas}(r)$ and $T_{\rm gas}(r)$
are reconstructed by using analytic 3D models with great functional
freedom, projecting them along a line of sight, and fitting the
projected models to observed profiles. The gas density model we use is
given by
\begin{equation}
\label{eq:density:model}
  \begin{split}
    n_p\,n_e = n_0^2\;\frac{(r/r_c)^{-\alpha}}{(1+r^2/r_c^2)^{3\beta-\alpha/2}}
                \;&
                \frac{1}{(1+r^\gamma/r_s{}^\gamma)^{\varepsilon/\gamma}}. 
  \end{split}
\end{equation}
This expression is a modification of the $\beta$-model
\citep{cavaliere_etal78} that allows independent modeling of changes
in the gas density slopes in cluster cores, outskirts, and
intermediate radii. \citet{vikhlinin_etal06} used an additional
$\beta$-model component with small core-radius to further increase
functional freedom in a very central region. We do not use this
feature and instead focus on modeling the surface brightness in a
region at $r>0.06\times \r500c$.

\begin{figure*}[t]  
  \vspace{-0.8cm} 
  \centerline{\includegraphics[height=7.8in]{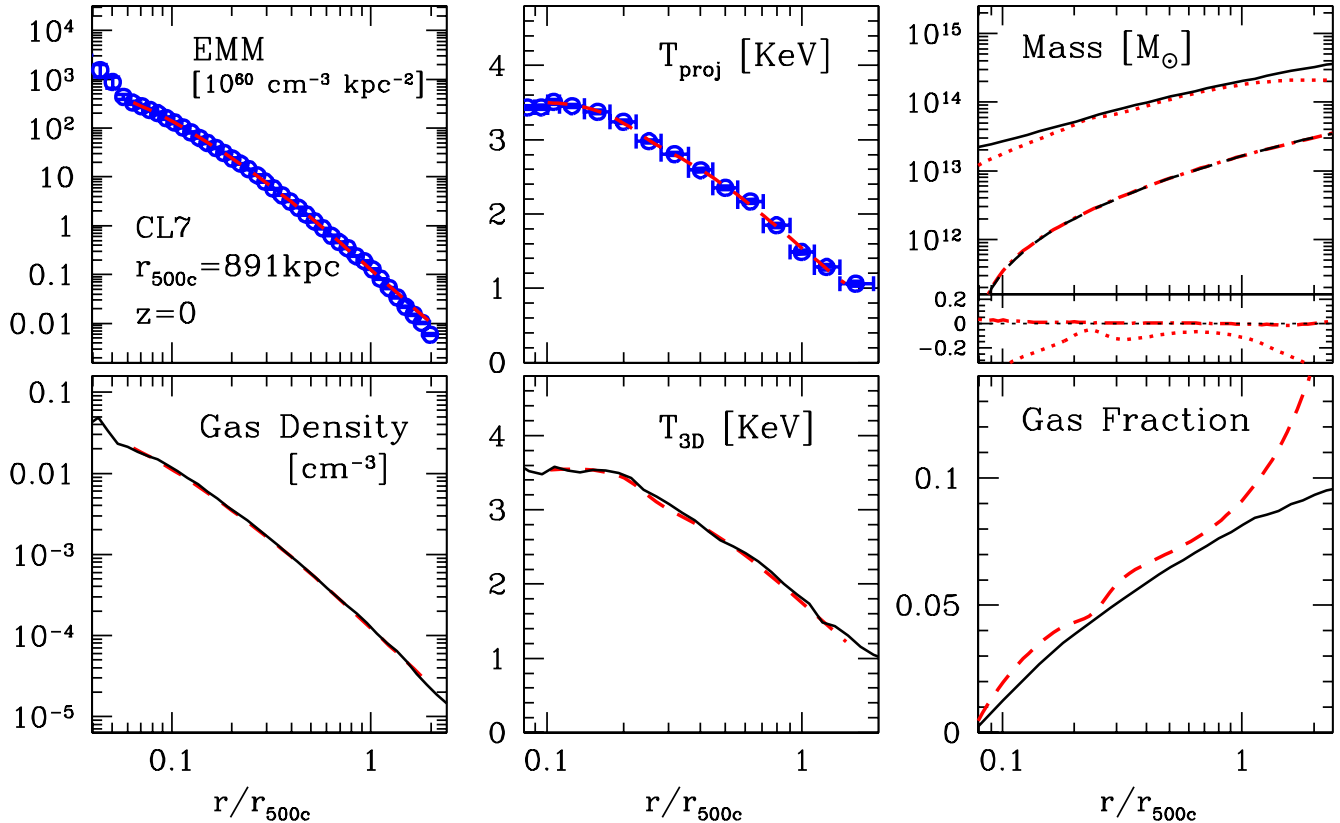}}
    \vspace{-7.5cm}
    \caption{The mock \emph{Chandra} analyses of a typical relaxed
      cluster, CL7, at $z=0$. The cluster has $\r500c=891$kpc,
      $\m500c=1.41\times 10^{14}h^{-1}\Msun$ and $\R2500c\approx
      0.5\r500c$.  The \emph{upper-left} panel shows the observed
      emission measure profile ($\approx$ surface brightness).  The
      measured projected temperature profile is shown in the
      \emph{upper-middle} panel.  In these panels, the dashed lines
      indicate best-fit projected profiles plotted for the radial range
      where the fits were made.  The \emph{lower-left} and
      \emph{-middle} panels show comparisons of the best-fit 3D model
      profiles and true gas density and temperature profiles.  Note that
      our model recovers well both the projected profiles and the actual
      3D profiles. In the \emph{upper-right} panel, we compare the
      derived and true $\Mtot$ (\emph{upper lines}) and $\Mgas$ profiles
      (\emph{lower lines}). The reconstructed $\Mgas$ profile
      \emph{dot-dashed} line) is accurate to a few percent in the entire
      radial range shown.  The hydrostatic $\Mtot$ estimate
      (\emph{dotted} line), on the other hand, is biased low by about
      5\%--10\% in the radial range, $[0.2,1.0]\r500c$.  The
      \emph{lower-right} shows that measured cumulative $\fgas$ is
      biased high by $\approx 10\%$ in the radial range of
      $[0.2,1.0]\r500c$ for this cluster, and it is primarily due to the
      bias in the hydrostatic mass estimate.}
\label{fig:CL7csf1_z0}
\end{figure*}

\begin{figure}[t]  
  \vspace{-0.57cm}
  \centerline{\epsfysize=3.8truein \epsffile{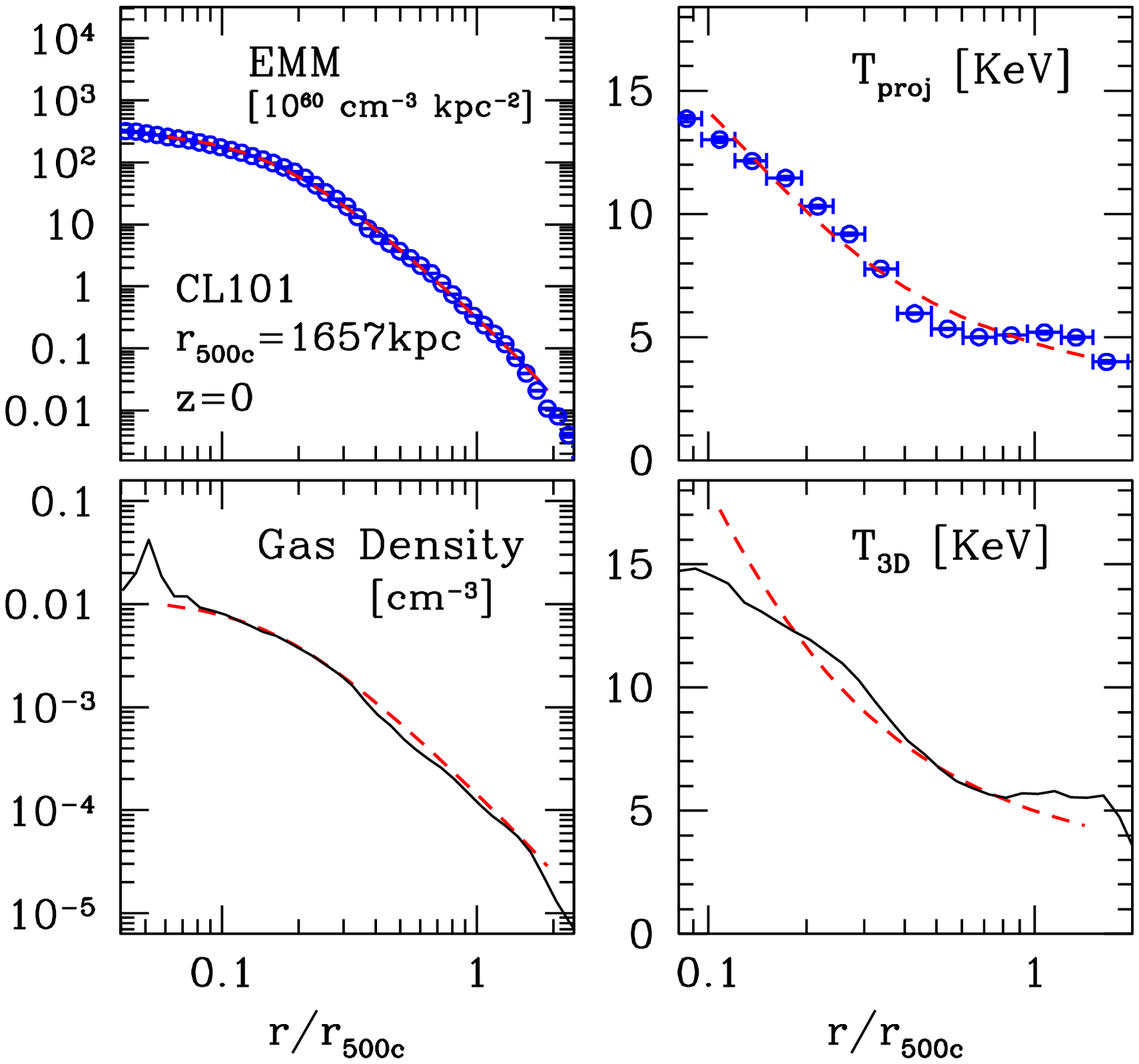}}
  \vspace{-0.8cm}
  \caption{ Same as Fig.~\ref{fig:CL7csf1_z0}, but for one of the
    unrelaxed clusters, CL101 at $z=0$. The emission measure and
    projected temperature profiles exhibit bumps and wiggles. The
    derived $\Mgas$ and $\Mtot$ profiles are thus less accurate than
    those in relaxed systems. However, the fits to the projected data
    are still good, highlighting the flexibility of our gas density and
    temperature models.}
\label{fig:CL101csf1_z0}
\end{figure}

The analytic expression of a 3D temperature profile was designed to
model general features of observed projected temperature profiles
\citep{vikhlinin_etal05c}.  The model consists of two terms:
\begin{equation}\label{eq:tprof}
  T_{\mathrm{3D}}(r) = t(r)\times t_{\text{cool}}(r). 
\end{equation}
The first term, $t(r)$, describes a temperature decline at large radii
by a broken power law profile with a transition region,
\begin{equation}\label{eq:tprof:main}
  t(r) = T_0\, \frac{(r/r_t)^{-a}}{(1+(r/r_t)^b)^{c/b}}. 
\end{equation}
The second term, $t_{\text{cool}}(r)$, is designed to model the
temperature decline in the central region affected by radiative cooling
\begin{equation}\label{eq:tprof:cool}
  t_{\text{cool}}(r) = (x+T_{\text{min}}/T_0)/(x+1), \quad
  x=(r/r_{\text{cool}})^{a_{\text{cool}}}. 
\end{equation}
To fit the observed projected temperature profile, the model is
projected along a line of sight. This projection requires proper
weighting of multiple temperature components, and we use the algorithm
that very accurately predicts a single-temperature fit to
multi-component spectra in a wide range of temperatures recently
proposed by \cite{vikhlinin_etal06b}. The observed projected
temperature profiles are fit in the radial range between $0.1\r500c$
and $r_{\rm out}={\rm max}(1.5\r500c,2/3r_{\rm max})$, where $r_{\rm
  max}$ is a half of the mock \emph{Chandra} image size.  The inner
radial range is set to exclude central components with unrealistic
X-ray properties found in most of our simulated clusters.  The exact
choice of $r_{\text{min}}$ is unimportant because we are primarily
interested in the cluster properties at large radii.  The choice of
the outer radius is motivated by a desire to ensure accurate
measurements of the temperature gradient and hence the $\Mtot$ at
$\r500c$ and that the fits in the cluster outskirts are not affected
by projection effects. 

Several examples of the observed surface brightness and projected
temperature profiles are shown in Figures~\ref{fig:CL7csf1_z0} and
\ref{fig:CL101csf1_z0} along with their best fit models. We find that
the models of eq.(\ref{eq:density:model}) and (\ref{eq:tprof}) usually
provide an excellent description of the data, except for very
irregular clusters such as that in Figure~\ref{fig:CL101csf1_z0} (its
X-ray image is shown in the right panel of
Figure~\ref{fig:images_z0}). 

Given 3D models for the gas density and temperature profiles, the
hydrostatic estimate of total cluster mass within a radius $r$ is
given by
\begin{equation}\label{eq:hydro}
  M(r) = -\frac{T(r)\, r}{\mu\, m_p G}\;
  \left(\frac{d\,\log \rho_g}{d\,\log r}+\frac{d\,\log T}{d\,\log r}\right),
\end{equation}
where $\mu=0.592$ is mean molecular weight of the fully ionized H-He
plasma used in the simulations. Given $M(r)$, we compute the total
mass at several overdensity levels, $\Delta$, relative to the critical
density at the cluster redshift, by solving equation
\begin{equation}\label{eq:overdensity}
  M_{\Delta}(r_\Delta) = \Delta\, 4/3\,\pi\,r_\Delta^3\,  \rho_c(z). 
\end{equation}
The ICM mass profile is obtained directly from the best-fit number density
profile (eq.\ref{eq:density:model}), $\rho_g = 1.236\,m_p\,(n_p n_e)^{1/2}$. 

Finally, we compute average temperatures for each cluster using
different weightings of the best-fit 3D temperature models and
observed projected profiles.  In this study, we consider several
definitions of average temperatures. The first is a X-ray spectral
temperature, $\Tspec$, a value derived from a single-temperature fit
to the integrated cluster spectrum excluding the core and detectable
small-scale clumps. We also compute a gas-mass-weighted temperature,
$\Tmg$, obtained by weighting of the reconstructed temperature profile
with the derived gas density profile.  We also consider an
emission-weighted temperature, $\Tew^{\rm SIM}$, computed by weighting
$T(r)$ with $n_pn_eT^{1/2}$; it is not directly observed but often
used in the theoretical work. Similarly, we compute a true
gas-mass-weighted temperature $\Tmg^{\rm SIM}$ from the simulations.
The superscript ``SIM'' indicates that the quantity is computed
directly from the 3D gas properties in the simulations without
excluding the {\it Chandra} detectable clumps, as opposed to from the
{\it Chandra} analysis fits, throughout this work. All these average
temperatures are computed excluding the central $0.15\,\r500c$ because
temperatures at these radii can be strongly affected by radiative
cooling and thus not directly related to the depth of the cluster
potential well. Also, because of the complex physics in the innermost
cluster regions the simulated ICM properties are probably not fully
realistic. The outer radius of integration is set to $\r500c$, which
approximately corresponds to the boundary within which the currently
observed cluster temperature profiles are reliable
\citep{vikhlinin_etal05c}.  However, we also consider a spectral
temperature $\Tspec[0.5\r500c]$ measured within a smaller aperture of
$0.15<r/\r500c<0.5$; this definition may be more practical than
$\Tspec$ integrated out to $\r500c$ for observations of limited
statistical quality or spatial coverage. Note that $\Tspec$ is
equivalent to $T_X$ defined in \citet{kravtsov_etal06}.

\section{Results}
\label{sec:results}


\subsection{Profiles of Relaxed clusters}
\label{sec:results_relaxed}

Figure~\ref{fig:CL7csf1_z0} shows the profile analysis of one of the
relaxed clusters, CL7, at $z=0$. The upper-left panel shows the
emission measure (EMM) profile constructed from the mock
\emph{Chandra} photon map.  The dashed line is a best-fit model in the
radial range between $0.06\r500c$ and $2\r500c$.  In this radial
range, the EMM profile is very smooth and the model provides an
excellent fit. We do not model an additional component associated with
dense and cool gas in the inner $\approx50$~kpc. Note, however, that
the $\Mgas(<\!\!50~{\rm kpc})$ is only a small fraction
($\approx 2\%$) of the $\Mgas(<\!\r500c)$. 

The upper-middle panel shows the projected temperature profile
obtained from the X-ray spectral fitting at each projected annulus.
Data points are plotted as a function of X-ray counts weighted radius,
and the error bars in the horizontal axis indicate a radial bin size
within which the X-ray spectrum was extracted.  The best-fit profile,
indicated by the dashed line, is a very good description of the data
in $0.1<r/\r500c<1.5$, the radial range where the fit was done.

In the lower-left and lower-middle panels, we compare the best-fit
model profiles from the mock \emph{Chandra} analysis to the true 3D
gas density and temperature profiles measured in the simulation.  For
the true profiles, we compute a density profile of hot, X-ray emitting
gas with $T>10^6$~K and a gas-mass-weighted temperature.  Note that a
volume-averaged temperature is quite similar to the gas-mass-weighted
average shown here.  For this cluster, both 3D gas density and
temperature profiles are recovered very accurately. 

The upper-right panel shows the comparisons of the $\Mtot$ and $\Mgas$
profiles derived from the mock \emph{Chandra} analysis and their
respective true profiles.  The comparisons show that the $\Mgas$
profile is unbiased, and it is recovered to better than 2\% in
$0.1<r/\r500c<2$, nearly the entire radial range where the EMM profile
was modeled. The $\Mtot$ derived from hydrostatic mass modeling, on
the other hand, is biased low by about 5\%--10\% in the radial range
of $0.2<r/\r500c<1$. The figure also shows that the measured $\fgas$
is biased high, on average, by about 10\% at all radii, mainly due to
the bias in the hydrostatic estimate of $\Mtot$. 

These results are representative of those for other clusters in the
relaxed cluster samples at both $z=0$ and $0.6$. The 3D gas density
and temperature profiles are recovered remarkably well, and the
$\Mgas$ profiles are accurate to \emph{a few per cent} at all radii.
The $\Mtot$, on the other hand, is typically biased low by about
5\%--20\% throughout the clusters.  One exception is the $\Mtot$
measurement for one the most relaxed clusters, CL104, at $z=0$, which
is nearly unbiased.  This system has not experienced a major merger
for several dynamical time ($\approx6$~Gyr).

\subsection{Profiles of Unrelaxed Clusters}
\label{sec:results_unrelaxed}

Accurate $\Mtot$ estimates in unrelaxed clusters cannot be obtained
from the hydrostatic equilibrium assumption. Reconstruction of 3D gas
mass and temperature profiles from projected X-ray data can be done
only approximately in such systems because of deviations from
spherical symmetry. However, many applications (e.g., cluster mass
function estimates) require analyses of full statistical samples that
include both relaxed and unrelaxed systems.  It is therefore important
to ask how well the observational analysis procedures can recover
properties of the ICM for unrelaxed systems.

Unrelaxed clusters are a diverse population with a wide range of
dynamical states. For the purpose of illustration, we show the profile
analysis for a cluster undergoing a major merger, CL101 at $z=0$
(Fig.\ref{fig:CL101csf1_z0}). This example highlights salient features
of unrelaxed systems. Compared to relaxed clusters, their EMM and
projected temperature profiles are less smooth and exhibit more
pronounced bumps and wiggles. These features are not recovered by our
smooth 3D model, even though the overall trends in the projected
temperature and especially in the EMM are reproduced reasonably well. 
Recovery of the 3D gas density and temperature profiles is much less
accurate than for relaxed systems.  In the case of CL101, for example,
the derived $\Mgas$ profile is biased high by about 5\%--10\% at all
radii, and the 3D temperature profile is biased low at $r>0.1\r500c$. 
If the hydrostatic equation is blindly applied to the best-fit models,
the derived $\Mtot$ is biased low by 25\% at $\r500c$. 

\begin{figure}[t]  
  \vspace{-1.0cm} 
  \hspace{1.5cm}
  \centerline{\epsfysize=5.8truein \epsffile{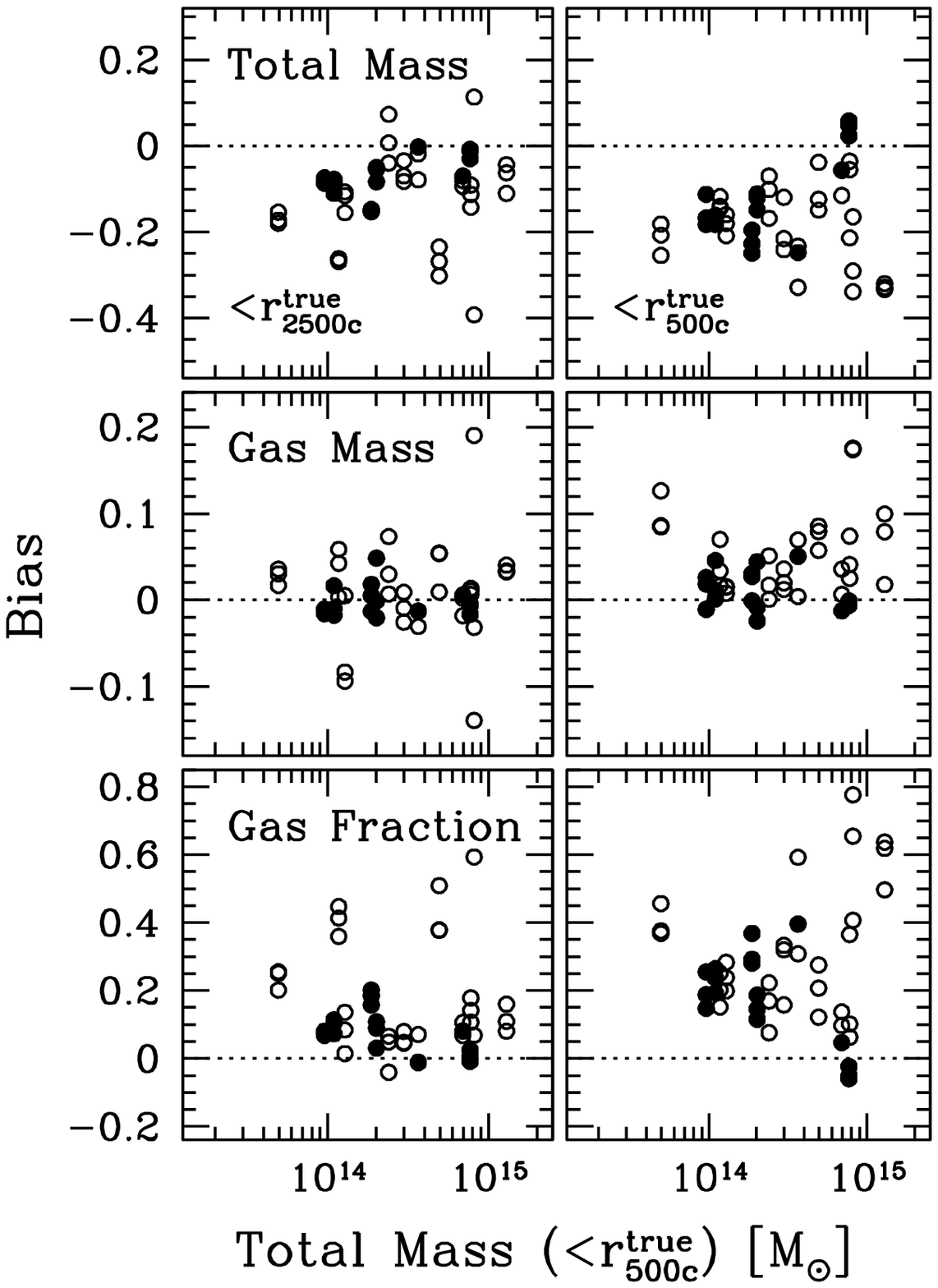}}
  \vspace{-1.5cm}
  \caption{ Bias in the $\Mtot$, $\Mgas$, and $\fgas$ measured within
    $\r500c^{\rm true}$ ({\it left}) and $\R2500c^{\rm true}$ ({\it
      right}) for simulated clusters at $z=0$.  The bias is defined as,
    e.g., $(M_{\rm est}-M_{\rm true})/M_{\rm true}$, where both $M_{\rm
      est}$ and $M_{\rm true}$ are at the same radius.  Each cluster is
    viewed along three orthogonal projections, and clusters with relaxed
    and unrelaxed morphologies are indicated with {\it filled} and {\it
      open} symbols, respectively.}
\label{fig:mbias_z0_true}
\end{figure}

\subsection{Gas Mass, Total Mass and Gas Mass Fractions}
\label{sec:masses}

In this section, we assess biases in the X-ray estimates of the enclosed
$\Mtot$, $\Mgas$, and $\fgas$.  The bias is defined as a fractional
difference between estimated mass, $M_{\rm est}$, derived from the mock
\emph{Chandra} analysis and true 3D mass, $M_{\rm true}$, measured
directly in simulations, ${\rm bias}=(M_{\rm est}-M_{\rm true})/M_{\rm
  true}$.  Figure~\ref{fig:mbias_z0_true} shows the results for all
simulated clusters at $z=0$ and at radii enclosing two different
overdensities: $\R2500c^{\rm true}$ and $\r500c^{\rm true}$.  Both
estimated ($M_{\rm est}$) and true masses ($M_{\rm true}$) are measured
in the same physical region enclosed within ``true'' radii ($r_{\rm
  true}$) measured directly from simulations.  Note also that the
$\Mgas$ is computed by excluding the mass enclosed within
0.075$\r500c^{\rm true}$\footnote{The choice is made to ensure that the
  central, dense and cool gas component, associated with the ISM of the
  central cluster galaxy, is excluded from the $\Mgas$ estimates of all
  clusters. This gas contributes $\lesssim 2\%$ of the total $\Mgas$.} 
(from both true and estimated masses), while the $\Mtot$ is obtained
without excluding the central region.  The $\fgas \equiv \Mgas/\Mtot$ is
therefore a fraction computed using these two quantities measured in a
slightly different radial range.  Table~\ref{tab:mbias1} also summarizes
average biases and scatter in the mass estimates at both $z=0$ and
$0.6$. 

The upper panels in Figure~\ref{fig:mbias_z0_true} show that the
hydrostatic mass estimate is typically biased low. It is
underestimated on average by 12\% at $\R2500c^{\rm true}$ and 16\% at
$\r500c^{\rm true}$ for all clusters at $z=0$.  The bias in the
hydrostatic mass is smaller in the inner region and it increases
toward cluster outskirts.  The bias is also smaller in relaxed systems
than in unrelaxed systems; for example, biases are 8\% and 15\% at
$\R2500c^{\rm true}$ for the relaxed and unrelaxed samples,
respectively. Note also that the scatter is very small inside
$\R2500c^{\rm true}$ of the relaxed clusters.  These results indicate
that the hydrostatic condition is best realized in the inner region of
relaxed clusters, while the deviation from the hydrostatic equilibrium
become more prominent in cluster outskirts and/or unrelaxed systems,
as expected. Similar results are obtained for clusters at high
redshift.

\begin{figure}[t]  
  \vspace{-0.55cm} 
  \centerline{\epsfysize=3.8truein \epsffile{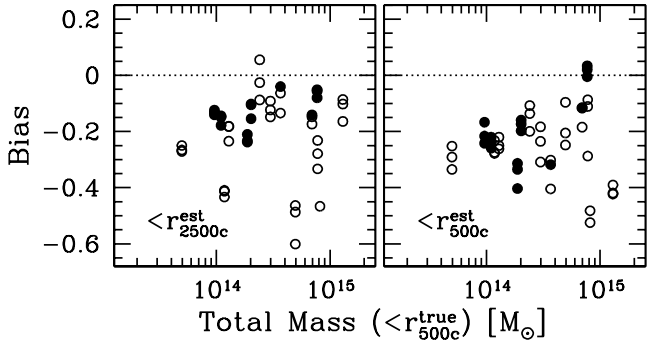}}
  \vspace{-4.3cm}
  \caption{ Same as the upper-panels in Fig.~\ref{fig:mbias_z0_true},
    except that the estimated mass ($M_{\rm est}$) is evaluated within
    the estimated virial radii ($r_{\rm est}$); hence, $M_{\rm est}$
    and $M_{\rm true}$ are measured in different physical regions.} 
\label{fig:mest_mtrue_all_z0}
\vspace{0.0cm}
\end{figure}

The middle panels in Figure~\ref{fig:mbias_z0_true} illustrate that
the X-ray gas mass determinations are remarkably accurate and robust. 
The \Mgas\ measurements are accurate to better than a few percent for 
both relaxed and unrelaxed clusters and independent of redshift.  
For the relaxed cluster samples, the accuracy of the \Mgas\ measurements 
is as good as 1\% at $\r500c^{\rm true}$ with 3\% $1~\sigma$ scatter. 
Although the effect is small ($\lesssim 6\%$), the \Mgas\ is biased 
high in the outskirts of the unrelaxed systems. 

The lower panels of Figure~\ref{fig:mbias_z0_true} show the biases in
measurements of the gas mass fractions.  Since the hydrostatic mass is
biased low, the derived \fgas{} are typically biased high.  The bias and
scatter in \fgas{} are especially large for the unrelaxed clusters,
since the biases in \Mgas{} and \Mtot{} are added constructively.  Both
bias and scatter are significantly reduced for relaxed clusters; for
example, $\fgas(<\!\R2500c^{\rm true})$ are accurate to about 10\% at
$z=0$ and $0.6$.  The biases become larger in cluster outskirts, and the
biases in $\fgas(<\!\r500c^{\rm true})$ are about 18\% at both low and
high redshifts. 

In practice, additional biases in the estimated cluster masses
($M_{\rm est}$) could arise from a bias in the estimation of a cluster
virial radius. Figure~\ref{fig:mest_mtrue_all_z0} shows the biases in
the estimated hydrostatic mass within the estimated virial radius,
$M_{\rm est}(<\!r_{\rm est})$, relative to the true cluster mass,
$M_{\rm true}(<\!r_{\rm true})$, measured in simulations.  The column
indicated as $M_{\rm tot}(<\!r_{\rm est})$ in Table~\ref{tab:mbias1}
summarizes average biases and scatter in the estimates of total
cluster masses computed this way. An underestimate of a cluster virial
radius results in the increased bias in the $\Mtot$ estimate and an
underestimate of the derived $M_{\rm gas}$ by a similar amount, while
leaving $f_{\rm gas}$ relatively unchanged.  It is these errors that
contribute to the differences in the mass-temperature relations
discussed in \S~\ref{sec:mt}.

A similar study of biases in the hydrostatic $\Mtot$ estimates has
been done recently by \citet{rasia_etal06}. The cosmological
simulations were performed using the Gadget-2 SPH code
\citep{2005MNRAS.364.1105S} and the mock X-ray data were reduced
closely following the procedure of \citet{2002A&A...391..841E}. Their
sample included 5 clusters, including two with relaxed morphology, for
which the comparison with our our results is most relevant. The mass
biases in the most comparable case (direct hydrostatic estimates and
reduced \emph{Chandra} background level) are $-30\%$ and $-28\%$ at
$\R2500c$, and $-15\%$ and $-32\%$ at $\r500c$. These values are on
the lower side of our distribution for relaxed clusters
(Fig.\ref{fig:mbias_z0_true}). In particular, none of our relaxed
clusters shows such strong biases at $r=\R2500c$. However, it is
difficult to compare our results directly. Not only the cosmological
simulation codes are quite different, but also the data analysis
algorithms are completely independent and significantly different.
Investigations of the sources of discrepancy will require cross-checks
(e.g., reduction of our mock data with the Rasia et al.{} pipeline)
and/or using a larger sample of Gadget-2 clusters.

\begin{figure}[t]  
  \vspace{-0.25cm} \hspace{0.0cm} 
  \centerline{\epsfysize=3.6truein \epsffile{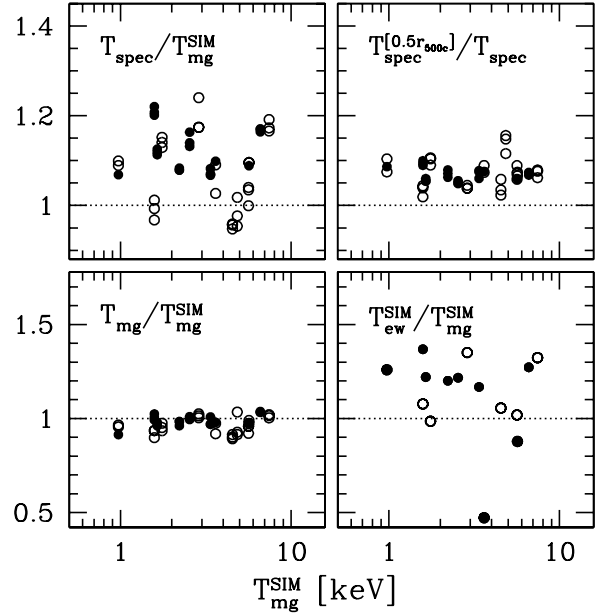}}
  \vspace{-0.8cm}
  \caption{ Comparisons of various average ICM temperatures for the
    $z=0$ sample. The temperature averages are defined in the text. The
    {\it filled} and {\it open} symbols indicate clusters with relaxed
    and unrelaxed morphology, respectively.  For each cluster results
    for the three projections are shown.}
\label{fig:tbias_z0_true}
\vspace{0.8cm}
\end{figure}

{
\def\m{\phantom{-}}
\begin{deluxetable*}{ccccccc}
\tablecaption{Bias in $\Mtot$, $\Mgas$ and $\fgas$ measurements.\label{tab:mbias1}}
\tablehead{
 & & & \multicolumn{4}{c}{Bias~$\pm$~Scatter} \\
\cline{4-7}\\[-1.7ex]
\colhead{Redshift} &
\colhead{Radial range} & 
\colhead{Sample} &
\colhead{~~$\Mtot(<\!r_{\rm true})$\tablenotemark{a}} &
\colhead{~~$\Mtot(<\!r_{\rm est})$\tablenotemark{a}} &
\colhead{~$\Mgas$\tablenotemark{b}} &
\colhead{$\fgas$\tablenotemark{c}}
}
\startdata
      &            & all       & $-0.121\pm0.136$  & $-0.222\pm0.181$&$\m0.006\pm0.047$ & $0.202\pm0.431$\\
$z=0$ & $<\R2500c$ & relaxed   & $-0.075\pm0.047$  & $-0.134\pm0.059$& $-0.003\pm0.017$ & $0.091\pm0.062$\\ 
      &            & unrelaxed & $-0.146\pm0.163$  & $-0.270\pm0.207$&$\m0.010\pm0.056$ & $0.268\pm0.525$\\ 
\hline                                                                          
      &            & all       & $-0.163\pm0.095$  & $-0.253\pm0.162$&$\m0.041\pm0.048$ & $0.264\pm0.184$\\
$z=0$ & $<\r500c$  & relaxed   & $-0.130\pm0.096$  & $-0.195\pm0.124$&$\m0.011\pm0.023$ & $0.176\pm0.136$\\ 
      &            & unrelaxed & $-0.182\pm0.091$  & $-0.285\pm0.173$&$\m0.058\pm0.051$ & $0.312\pm0.190$\\ 
\hline                                                                        
       &           & all       & $-0.122\pm0.116$  & $-0.245\pm0.151$&$\m0.010\pm0.045$ & $0.175\pm0.216$\\
$z=0.6$& $<\R2500c$& relaxed   & $-0.100\pm0.024$  & $-0.178\pm0.038$&$\m0.000\pm0.022$ & $0.112\pm0.038$\\ 
       &           & unrelaxed & $-0.135\pm0.145$  & $-0.286\pm0.178$&$\m0.016\pm0.054$ & $0.213\pm0.266$\\
\hline                                                                          
       &           & all       & $-0.089\pm0.206$  & $-0.144\pm0.306$&$\m0.024\pm0.072$ & $0.180\pm0.272$\\
$z=0.6$& $<\r500c$ & relaxed   & $-0.152\pm0.069$  & $-0.214\pm0.079$&$\m0.007\pm0.029$ & $0.196\pm0.111$\\ 
       &           & unrelaxed & $-0.052\pm0.250$  & $-0.102\pm0.379$&$\m0.034\pm0.087$ & $0.171\pm0.335$
\enddata
\tablenotetext{a}{Estimated masses, $M_{\rm est}$, enclosed within
  $r_{\rm true}$ and $r_{\rm est}$ are compared with true cluster
  masses, $M_{\rm true}(<\!r_{\rm true})$, measured in simulations.} 
\tablenotetext{b}{Enclosed $\Mgas$ within $\r500c$ and $\R2500c$
  excluding the mass enclosed within $0.075\r500c$.} 
\tablenotetext{c}{A fraction computed using above two quantities.} 
\end{deluxetable*}
}

\begin{deluxetable*}{cccccc}
\tablecaption{Ratio of average cluster gas temperatures\label{tab:tbias1}}
\tablehead{
 & & \multicolumn{4}{c}{Average~$\pm$~Scatter} \\
\cline{3-6}\\[-1.7ex]
\colhead{Redshift} &
\colhead{Sample} & 
\colhead{$\Tmg/\Tmg^{\rm SIM}$} &
\colhead{$\Tspec/\Tmg^{\rm SIM}$} &
\colhead{$\Tew^{\rm SIM}/\Tmg^{\rm SIM}$} &
\colhead{$\Tspec^{[0.5\r500c]}/\Tspec$} 
}
\startdata
      & all       & $0.973\pm0.040$ &  $1.095\pm0.078$ & $1.078\pm0.283$ & $1.072\pm0.027$ \\
$z=0$ & relaxed   & $0.990\pm0.029$ &  $1.126\pm0.049$ & $1.188\pm0.192$ & $1.070\pm0.015$ \\ 
      & unrelaxed & $0.960\pm0.042$ &  $1.070\pm0.087$ & $0.992\pm0.315$ & $1.073\pm0.034$ \\  
\hline \hline                                                                             
       & all      & $0.985\pm0.047$ &  $1.080\pm0.075$ & $1.511\pm0.619$ & $1.044\pm0.019$ \\   
$z=0.6$& relaxed  & $0.985\pm0.029$ &  $1.110\pm0.056$ & $1.725\pm0.789$ & $1.048\pm0.007$ \\    
       & unrelaxed& $0.985\pm0.054$ &  $1.066\pm0.079$ & $1.404\pm0.459$ & $1.042\pm0.022$ 
\enddata
\end{deluxetable*}

\subsection{Average Temperatures}
\label{sec:tave}

Average cluster temperature, $\Tave$, is another important ICM
diagnostic and a key observable for cosmological application. Since
the ICM is not isothermal within a cluster, the definition of $\Tave$
is not unique. The differences among various definitions should be
calibrated and taken into account.  Figure~\ref{fig:tbias_z0_true}
compares different $\Tave$ obtained from the mock \emph{Chandra}
analysis, $\Tspec$, $\Tspec^{[0.5\r500c]}$, and $\Tmg$, as well as
those measured directly from the 3D properties of gas in simulations,
$\Tew^{\rm SIM}$ and $\TmgSIM$, for all clusters at $z=0$ and $0.6$
(see \S\,\ref{sec:MockChandra2}). The gas with $T<0.086$~keV
($10^6$~K) is excluded from all calculations of average temperatures,
since it does not contribute to the X-ray flux. Note also that
detectable small clumps are excluded from the mock Chandra analysis
and hence do not affect the determination of $\Tave$. Average ratios
of different temperature definitions are summarized in
Table~\ref{tab:tbias1}.

The best agreement is between the true and X-ray derived gas
mass-weighted temperatures. Other definitions show a constant offset
relative to $\TmgSIM$ with some scatter around the mean.  For the
relaxed clusters at $z=0$, we find $\Tew^{\rm
  SIM}:\Tspec:\Tmg:\TmgSIM=1.19:1.13:0.99:1$. The ratios are slightly
different for the non-relaxed clusters, $\Tew^{\rm
  SIM}:\Tspec:\Tmg:\TmgSIM=0.99:1.07:0.96:1$. The $\Tspec$ is higher
than $\Tmg$ because the former is dominated by the inner, hotter
region (see examples of the temperature profiles in
Fig.~\ref{fig:CL7csf1_z0} and~\ref{fig:CL101csf1_z0}). A tight
correlation exits between $\Tspec$ and $\TmgSIM$ with the
object-to-object scatter as small as 5\% for the relaxed clusters and
9\% even for the unrelaxed systems.  Although computed directly from
the simulation outputs, $\Tew^{\rm SIM}$ shows a much poorer
correlation with $\TmgSIM$ than any of the observationally derived
average temperatures.  The scatter in the temperature ratios is
generally higher for non-relaxed systems.

The X-ray spectral temperature, $\Tspec$, is of particular interest to
X-ray observers, because it is the most easily measured spectral
characteristic.  Our study shows that $\Tspec$ is related to $\Tmg$ by
a constant factor, $\Tspec/\Tmg\approx1.14$, for the relaxed clusters
and a slightly smaller factor, $\Tspec/\Tmg=1.12$, in the unrelaxed
systems. For each sub-sample, the ratio $\Tspec/\Tmg$ and its scatter
are nearly identical at $z=0$ and $0.6$ and do not show any sign of
evolution with redshift. Similar results are obtained for
$\Tspec^{[0.5\r500c]}$, which is related to $\Tspec$ by a constant
factor, $\Tspec^{[0.5\r500c]}/\Tspec\approx1.04-1.07$ with little
redshift trend and scatter, indicating that $\Tspec^{[0.5\r500c]}$ can
be used as a reliable substitute for $\Tspec$.  Finally, the ratio of
$\Tspec/\Tmg$ and its scatter from cosmological simulations is in good
agreement with the \emph{Chandra} results for low-redshift relaxed
clusters \citep{vikhlinin_etal06}, where the average ratio in
[0.15,1]$\r500c$ is 1.15 and scatter is about 0.08.

\subsection{Implications for the Mass-Temperature relation}
\label{sec:mt}

The results in \S~\ref{sec:masses}--\ref{sec:tave} have direct
implications for the mass-temperature ($M-T$) relation. The
normalization of the $M-T$ relation, $M_5$, can be defined as
\begin{equation} \label{eq:mt}
E(z)\, M_{500c} = M_{5} \left( \frac{\langle T \rangle}{5~{\rm keV}} \right)^{3/2}
\end{equation}
where $E(z)=(\Omega_M(1+z)^{3}+\Omega_{\Lambda})^{1/2}$ for a flat
universe with a cosmological constant. Table~\ref{tab:mt} lists the
best-fit values of $M_{5}$ and its uncertainties (arising because of the
finite sample size) for different temperature averages \citep[see also
Table 2 and Fig.~\ref{fig:tbias_z0_true} in][]{kravtsov_etal06}.  The
values of $M_5$ are computed using both the true mass, $M_{500c}^{\rm
  true}(<\!r_{500c}^{\rm true})$, measured in simulations and the
estimated mass, $M_{500c}^{\rm est}(<\!r_{500c}^{\rm est})$, derived
from the mock analysis. For $M_{500c}^{\rm true}$, the $M_{5}$ is higher
by about 10\% than their observed counterparts at both low and
high-redshifts: $M_{5}=3.06\pm0.16$ ($\Tspec$) and $3.64\pm0.18$
($\Tmg$) at $z\sim0$ \citep{vikhlinin_etal06} and $M_{5}=3.36\pm0.32$
($\Tspec$) at $z\sim0.5$ \citep{kotov_vikhlinin_06}.\footnote{To compare
  the M-T relations from simulations and observations in the same radial
  range (0.15\r500c-\r500c), we applied the correction factors of
  $\langle T(0.15\r500c-\r500c)/T(70~{\rm kpc}-\r500c) \rangle = 0.97$
  (for $\Tspec$) and $0.94$ (for $\Tmg$) to the observed relations.}
The $M_{5}$ based on $M_{500c}^{\rm est}$, on the other hand, is lower
by about 10\%--15\%.  In other words, the observed relations lie between
the simulations derived relations based on the true and estimated
masses.

\section{Discussion and conclusions}
\label{sec:discussion}

We present mock \emph{Chandra} analyses of cosmological cluster
simulations and assess X-ray measurements of galaxy cluster
properties.  To test observational X-ray procedures, we construct mock
{\it Chandra} images of the simulated clusters and derive properties
of X-ray clusters using a model and procedure essentially identical to
those used in real data analysis. The sample includes 16 clusters
spanning a broad mass range ($5\times 10^{13}-2\times
10^{15}\,h^{-1}\,M_{\odot}$) simulated in the $\Lambda$CDM cosmology
with high spatial resolution and including various physical processes
of galaxy formation, such as radiative cooling, star formation and
other processes accompanying galaxy formation.  We analyze the
simulated clusters at $z=0$ and their most massive progenitors at
$z=0.6$.  To test the sensitivity of the X-ray analysis to the cluster
dynamical state and substructure, we also distinguish unrelaxed
clusters from relaxed systems based on the overall structural
morphology of the \emph{Chandra} images, as usually done to classify
observed clusters. Below we discuss and summarize our main findings
and conclusions.

\begin{deluxetable}{cccccc}
\tablecaption{Normalizations of the $M_{500c}-T$ relation\label{tab:mt}}
\tablehead{
 & & & \multicolumn{3}{c}{$M_5$, $10^{14}\,h^{-1}\,M_\odot$}\\
\cline{4-6}\\[-1.7ex]
\colhead{Mass\tablenotemark{a}} &
\colhead{Redshift} &
\colhead{Sample} & 
\colhead{$\Tmg$} &
\colhead{$\Tspec$} &
\colhead{$\Tspec^{[0.5\r500c]}$} 
}
\startdata
                     &       & all       & $4.46\pm0.18$ & $3.78\pm0.21$ & $3.40\pm0.17$ \\
$M^{\rm true}_{500c}$&$z=0$  & relaxed   & $4.02\pm0.10$ & $3.32\pm0.10$ & $3.00\pm0.09$ \\ 
                     &       & unrelaxed & $4.80\pm0.19$ & $4.14\pm0.22$ & $3.71\pm0.18$ \\  
\hline                                                                                  
                     &       & all       & $4.65\pm0.17$ & $4.07\pm0.17$ & $3.81\pm0.16$ \\     
$M^{\rm true}_{500c}$&$z=0.6$& relaxed   & $4.50\pm0.12$ & $3.77\pm0.12$ & $3.51\pm0.12$ \\      
                     &       & unrelaxed & $4.73\pm0.18$ & $4.22\pm0.17$ & $3.96\pm0.16$ \\ 
\hline
\hline                                                                           
$M^{\rm est}_{500c}$ &$z=0$  & relaxed   & $3.14\pm0.03$ & $2.59\pm0.03$ & $2.35\pm0.04$ \\[1mm]
\hline                                                                                  
$M^{\rm est}_{500c}$  &$z=0.6$& relaxed   & $3.55\pm0.07$ & $2.97\pm0.06$ & $2.77\pm0.06$\\[-2mm]      
\enddata
\tablenotetext{a}{$M^{\rm true}_{500c}$ is the true $\Mtot$ of a
  cluster enclosed within the true $\r500c^{\rm true}$ measured
  directly in simulations; $M^{\rm est}_{500c}$ is the estimated
  hydrostatic mass enclosed within the estimated $\r500c^{\rm est}$
  from the mock data analysis.} 
\end{deluxetable}

\subsection{Reconstruction of 3D ICM Properties}

Derivation of the cluster 3D properties from projected X-ray data
usually relies on simplifying assumptions such as spherical symmetry
of a cluster or small-sale uniformity of the ICM. Additional
complications arise from a non-ideal response of X-ray telescopes and
limited statistical quality of data. This leads to several effects
which are hard to reproduce analytically without creating mock X-ray
data from realistic cluster simulations. For example, clumps of colder
gas associated with massive subhalos can be detected and masked out
from a further analysis while smaller subhalos remain undetected and
bias projected temperature and X-ray brightness measurements
\citep{mazzotta_etal04,vikhlinin_etal06b}. One of the goals of our
work is to include all such effects as completely as possible in
generation of the mock X-ray data and thus to test how they bias X-ray
measurements and recovery of 3D ICM profiles. 

Reassuringly, we find that the ICM models and analysis method
presented in \citet{vikhlinin_etal06} provide a good description of
emission measure and projected temperature profiles of the ICM.  For
relaxed clusters, these models recover the 3D gas density,
temperature, and gas mass profiles with the accuracy of {\it a few
  percent} at $0.1<r/\r500c<2.0$.  The models are flexible enough to
describe also non-relaxed systems and provide reasonably accurate gas
mass and temperature measurements. A practical implication of these
results is that projection effects, cluster substructure, and
deviations from spherical symmetry do not strongly affect
reconstruction of 3D ICM density, temperature, pressure, and entropy
profiles from X-ray data.  Therefore, observational results on ICM
profiles can be directly compared to the properties of simulated
clusters without a need for detailed mock X-ray analyses. Such a
comparison for our sample will be presented in the future paper
\citep{nagai_etal06c}.

The total ICM mass is measured quite accurately in all our simulated
clusters. Results are most accurate for relaxed systems: biases and
scatter in \Mgas\ measurements are accurate to better than $\sim 3\%$
percent, independent of a cluster redshift or a radial range within in
which the \Mgas\ is measured.  For unrelaxed systems, the scatter is
larger, as expected \citep{mathiesen_etal99}, but the bias is still
small ($\lesssim 10\%$).

Average temperature is an important diagnostic of properties and
physical processes of the ICM and a key observable for cosmological
applications. Since clusters are not isothermal, the definition of the
$\Tave$ is not unique nor can $\langle T\rangle$ be completely
accurately derived from data. Mock X-ray analysis is therefore
required to study the relation among different $\langle T\rangle$
definitions used in observational and theoretical studies.  Our
analysis is focused on the comparisons of the true gas-mass-weighted
average temperature, $\TmgSIM$, and the X-ray spectroscopic $\Tspec$
or gas-mass-weighted $\Tmg$ temperatures derived from the mock data.
We find that these temperatures are not identical, but the difference
is mostly a constant factor with a relatively small object-to-object
scatter.  For example, $\Tspec /\Tmg \approx 1.14\pm0.05$ for relaxed
clusters, while for unrelaxed objects this ratio is $\Tspec/\Tmg
\approx 1.12\pm0.09$.  All the $\Tave$ definitions we tested are
therefore equivalent, but the difference should be kept in mind when,
e.g., observed $M-T$ relations are compared with results of
simulations.  A notable exception is the ``emission-weighted''
temperature defined as $\Tew = \int T\, \rho^2\,T^{1/2}\,dV \big/ \int
\rho^2\,T^{1/2}\,dV$
\citep{bryan_etal98,frenk_etal99,muanwong_etal01}. Even computed
directly from the simulation outputs, this quantity shows a much
poorer correlation with $\TmgSIM$ than any of the observationally
derived average temperatures (Fig.\ref{fig:tbias_z0_true}). Our
results thus support the suggestion of \citet{mazzotta_etal04} that
$\Tew$ should not be used as a measure of the average cluster
temperature.

\subsection{Accuracy of Hydrostatic Mass Estimates}
\label{sec:accur-hydr-total}

One of the major applications of the gas density and temperature
profile measurements from the X-ray data is the estimate of a
gravitationally bound mass of a cluster assuming that the ICM is in
hydrostatic equilibrium in the cluster potential. The X-ray method is
by far the most precise way to estimate $\Mtot$ in individual objects
at large radii. This limits our ability to test biases in the X-ray
derived $\Mtot$ estimates through comparisons with independent
observational techniques such as weak lensing. The analysis of our
mock X-ray data, however, provides important clues. 

We find that the hydrostatic mass estimates from our mock data are
biased low by about 5\%--20\% throughout the virial region
(Fig.\ref{fig:mbias_z0_true}), even in clusters that are identified as
relaxed by their X-ray morphology. Such biases seem to be too high to
be attributed to inaccuracies in 3D gas density and temperature
profile reconstruction. A more plausible explanation is a departure of
the ICM from a hydrostatic state. Indeed, an additional pressure
component due to residual random bulk motions of the ICM in our
simulated clusters completely explains the bias in the hydrostatic
mass estimates \citep[this topic will be discussed in greater detail
in][]{lau_etal06}.  Similar results have been also obtained by
independent simulations performed with different codes and
implementation of physical processes
\citep{rasia_etal06,dolag_etal05b}. The $\Mtot$ bias derived in the
mock X-ray analysis of our relaxed clusters increases towards their
outskirts. The average levels are $-8\%$ at $r=\R2500c$ and $-13\%$ at
$r=\r500c$, and quickly become very large beyond $\r500c$
(Fig.\ref{fig:CL7csf1_z0}), which qualitatively mimic the radial
dependence of turbulent pressure observed in previous simulations
\citep[e.g.,][]{evrard_etal96}. In unrelaxed clusters, not only the
departures from hydrostatic equilibrium are strong but also the
density and temperature gradients are not measured accurately
(Fig.\ref{fig:CL101csf1_z0}) resulting in $\Mtot$ biases of 20\% or
more at all radii. 

It is unclear to what degree the level of ICM turbulence found in
these simulations applies to real clusters because most of the codes
model ICM as an ideal inviscid fluid with some amount of numerical
viscosity.  Further progress in this area using cosmological
simulations will have to rely on better understanding and modeling of
viscosity of the weakly magnetized plasma
\citep{ruszkowski_etal04a,ruszkowski_etal04b,dolag_etal05b,sijacki_etal06}. 
Direct measurements of the ICM turbulence through Doppler broadening
of emission lines \citep{inogamov_etal03} must await a launch of X-ray
calorimeters, although some clues could be provided also by
observations of X-ray brightness fluctuations
\citep{schuecker_etal04}. Keeping these uncertainties in mind, we can
treat the $\Mtot$ biases quoted above as upper limits. 

Cosmic rays and magnetic fields may also provide non-thermal pressure
support which is not accounted for in the hydrostatic estimates nor
included in our simulations. In addition, temperatures of electron and
ion components of the ICM are not necessarily the same, because strong
shocks heat primarily ions and the collisional electron-ion
equilibration time is long in the cluster outskirts
\citep{markevitch_etal96,fox_etal97}. If $T_e$ (derived from X-ray
spectra) is indeed lower that $T_i$ (unobserved), the hydrostatic mass
estimates are biased low. We note, however, that at least in one case
there is evidence for fast equilibration of the electron and ion
temperatures in a post-shock region of the ICM \citep{markevitch05}. 

While direct measurements or theoretical modeling of the non-thermal
pressure support may be difficult, it can be constrained by comparisons
of the hydrostatic mass estimates with those derived from gravitational
lensing for the same objects. An important caveat is that lensing
measures the mass in the projected aperture while the mass within the
sphere of the same radius is rather uncertain because of projection
effects.  \citet{metzler_etal01} estimate that $M_{500c}$ cannot be
determined from weak lensing with better than $\pm$30\% uncertainty. Any
comparisons of X-ray hydrostatic and weak lensing masses must be done in
an average sense, e.g.{} through derived normalizations of the
correlation between the $\Mtot$ and a low-scatter mass proxy such as
$Y_X$ \citep{kravtsov_etal06} or $\Tspec$. 

To summarize, our results indicate that an X-ray analysis of relaxed
clusters correctly recovers gas density and \emph{thermal electron}
pressure gradients. The simulations provide an upper limit (5\%--20\%
depending on radius) for the mass bias due to turbulent motions.  The
contribution of other pressure components will have to be constrained
by future observations.

\subsection{Implications for Cluster-Based Cosmological Tests}

\subsubsection{Tests Based on $\fgas$}

X-ray measurements of the baryonic mass fraction in massive clusters
can be used to constrain $\Omega_M$ or $h$ \citep{white_etal93} and
also provide a potential standard candle
\citep{sasaki96,pen97,allen_etal04}.  The major observational
ingredient in these tests is a measurement of the hot gas mass 
fraction within a sufficiently large radius. Our analysis provides 
clues for biases in $\fgas$ measurements from high-quality 
\emph{Chandra} data. We find that the $\fgas$ determinations 
are biased high, primarily because the
hydrostatic method underestimates the total cluster mass. As was
discussed above, the bias in the hydrostatic estimates is related to
the physical processes in the ICM (e.g., turbulence, cosmic ray and
magnetic pressures etc.) and not to the deficiencies of the X-ray
analysis. Further progress in confirming validity of $\fgas$
measurements will have to rely on independent determinations of 
cluster total mass (see discussion in \S~\ref{sec:accur-hydr-total}). Some
sources of the bias (e.g., turbulence) could be minimized if we focus
only on relaxed systems. We find that in such systems, both bias and
scatter in $\fgas$ can be controlled to within about 5\%--8\% at
$\R2500c$ and 15\%--20\% at $\r500c$, which can be considered the
upper limits on the $\fgas$ measurement bias caused by turbulence (see
\S~\ref{sec:accur-hydr-total}). 

Note that if the ICM turbulence and other processes discussed in
\S~\ref{sec:accur-hydr-total} do play a role in biasing the
hydrostatic mass estimates, the bias can potentially be
redshift-dependent since the merger rate increases at higher $z$
\citep[e.g.][]{gottloeber_etal01}. We do not find any detectable
redshift dependence of the bias in the $\fgas$ measurements in relaxed
systems. However, our upper limits, $\sim\! 5\%$, are weaker than the
accuracy required, e.g., for using $\fgas$ as a standard ruler in the
dark energy studies \citep{allen_etal04}. Mock analysis of much larger
cluster samples will be required to properly address this issue.

\subsubsection{Tests Based on the Cluster Mass Function}

Another group of cosmological tests are those based on measurements of
the cluster mass function. The relevant issue for our work is how well
various mass proxies can be measured. A related question is how well
the cluster mass vs.{} proxy relation can be calibrated by the X-ray 
analysis, which was discussed in \S~\ref{sec:accur-hydr-total}. 

We did not discuss the easiest X-ray mass proxy, the total luminosity,
because it is also the least accurate and least reliably reproduced by
numerical simulations. The next, in terms of being easily measured, is
the average ICM temperature. The practical averages to use are
$\Tspec$ or $\TspecIV$\footnote{Other definitions such as $\Tmg$
  require measurements of the temperature \emph{profiles} which is
  feasible only for a few best-observed clusters.}. These definitions
are freely interchangeable since the ratio $\TspecIV/\Tspec$ is nearly
the same in all clusters (Fig.\ref{fig:tbias_z0_true},
Table~\ref{tab:tbias1}).  One disadvantage of the spectral temperature
as a mass proxy is that its measurement is somewhat sensitive to the
cluster dynamical state and small-scale substructure. This sensitivity
arises because $\Tspec$ is rather sensitive to the presence of
lower-temperature components. This effect contributes to the
systematic difference in the normalizations of the $M-T$ relation
between relaxed and non-relaxed clusters and also slight deviations
from the self-similar evolution (Table~\ref{tab:mt}).  Some
implications of this effect for the cosmological measurements are
discussed by \citet{rasia_etal05}. The prospects for calibrating the
$\Tspec$ measurement biases by cosmological numerical simulations are
mixed. The effects related to the cluster dynamics, such as merger
activity and shocks etc.{}, are treated sufficiently accurately.
However, the biases caused by the presence of undetected cold clumps
will require a more realistic modeling of the ICM cooling and
associated feedback.

More accurate X-ray proxies for cluster total mass is the mass of the
hot gas, $\Mgas$, and the ``X-ray $Y$-parameter'',
$Y_X=\Mgas\times\Tspec$ \citep{kravtsov_etal06}. The properties of
$Y_X$ (this is the best X-ray mass proxy known) are extensively
discussed elsewhere \citep{kravtsov_etal06}, so we concentrate on
$\Mgas$ here. The $\Mgas$ is measured very reliably. The average bias
is $<5\%$ with only a few strong outliers
(Fig.\ref{fig:mbias_z0_true}) and there is no detectable redshift
dependence. A similar accuracy should in principle be achieved for the
$\Mtot$ estimates since to the first order, $\Mtot\propto\Mgas$. Note
that the biases in $\Mgas$ measurements are mostly due to large-scale
deviations of the cluster body from spherical symmetry. This effect
should be reliably treated in the current generation cosmological
simulations, and so the accuracy of the $\Mgas$-based total mass
estimates can be improved yet further. All we need is a larger sample
of simulated clusters. A major obstacle, however, is the uncertainties
in the mass fraction of cold baryons (stars and molecular gas). This
fraction can be substantial, as indicated by our simulations
\citep{kravtsov_etal05} and significant deviations of the observed
$\fgas$ in clusters from the cosmic mean
\citep[e.g.,][]{ettori03,vikhlinin_etal06}. Further progress in this
area requires significant improvements in the reliability of the
stellar mass measurements and treatment of the star formation in the
simulations. We will further discuss these issues in the forthcoming
papers.

\smallskip

To briefly summarize our results, we find that the analysis of deep
X-ray observations of galaxy clusters can reliably recover its
intended targets: the distribution of density and temperature of the
hot ICM. The accuracy of estimating the total cluster mass from these
parameters is limited by additional physical processes in the ICM,
such as turbulence, magnetic and cosmic ray pressure, and possible
departures from the electron-ion equilibrium. Our work provides a
realistic estimate of the effects of turbulence, which are found to
cause underestimation of $\Mtot$ by 5\%--20\% in clusters that can be
visually classified as relaxed (this can be considered as an upper
limit for the bias if the ICM viscosity is non-negligible).
Constraining the role of other effects will probably have to rely on
accurate independent mass measurements (e.g., by weak lensing) in a
large, complete cluster sample.

\acknowledgements We thank Elena Rasia for a careful reading of this
manuscript. DN is supported by the Sherman Fairchild Postdoctoral
Fellowship at Caltech. AV is supported by the NASA grant NAG5-9217 and
contract NAS8-39073.  AVK is supported by the National Science
Foundation (NSF) under grants No.  AST-0239759 and AST-0507666, by
NASA through grant NAG5-13274, and by the Kavli Institute for
Cosmological Physics at the University of Chicago.  The cosmological
simulations used in this study were performed on the IBM RS/6000 SP4
system ({\tt copper}) at the National Center for Supercomputing
Applications (NCSA).


\end{document}